\def\o{\over}
\def\Ar{\rightarrow}
\def\bar{\overline}

\def\d{\delta}
\def\a{\alpha}

\def\n{\nu}
\def\m{\mu}
\def\k{\kappa}
\def\e{\epsilon}

\def\th{\theta}

\def\t{\tilde}
\def\bar{\overline}
\def\l{\lambda}
\def\G{{\rm GeV}}

\def\eV{{\rm eV}}
\documentstyle [12pt]{article}
\begin{document}
\baselineskip=24.5pt
\setcounter{page}{1}
\thispagestyle{empty}
\topskip 0.5  cm
\vspace{1 cm}
\centerline{\Large\bf  Neutrino mass matrix with U(2) flavor symmetry}
\centerline{\Large\bf    and neutrino oscillations}
\vskip 1.5 cm
\centerline{{\bf Morimitsu TANIMOTO}
  \footnote{E-mail address: tanimoto@edserv.ed.ehime-u.ac.jp}}
\vskip 0.8 cm
 \centerline{ \it{Science Education Laboratory, Ehime University, 
 790-77 Matsuyama, JAPAN}}
\vskip 2 cm
\centerline{\bf ABSTRACT}\par

 The three neutrino mass matrices in the $SU(5)\times U(2)$  model are studied
  focusing on the neutrino oscillation experiments. 
 The atmospheric neutrino anomaly could be explained by the large 
  $\nu_{\mu} - \nu_{\tau}$ oscillation.
   The long baseline experiments are expected to detect signatures of the 
  neutrino oscillation even if the atmospheric neutrino anomaly is not due to
  the neutrino oscillation.
  However,  the model cannot solve the solar neutrino deficit
  while it could be reconciled with the  LSND data.

\vskip 0.5 cm
 
 \newpage
\topskip 0.  cm
\voffset = -1 cm
\hoffset = -.50 in
   The  U(2) flavor symmetry \cite{U21,LNS} is an interesting candidate
   for beyond SM. This flavor symmetry should be more precisely tested by 
   future experiments \cite{U22}.
	The significant test  will be possible
    in the neutrino sector as well as in the quark sector.
  
 The extension of the $U(2)$ model to neutrino masses and mixings has been
presented by
  Carone and Hall \cite{CaHa}.  They have found that a simple modification
is required 
  for the flavor symmetry breaking pattern
  if the light three neutrino masses are obtained by the see-saw mechanism.
  This fact suggests  that the $U(2)$ symmetry will be tested seriously
   in  neutrino oscillations.
 The purpose of our paper is to present the systematic analyses in order to
 clarify the phenomenological implications of the $SU(5)\times U(2)$  model
   in  neutrino oscillations.


  In the $U(2)$ flavor symmetry \cite{U21} the lighter two generations
transform as 
	 a doublet and the third generation as a singlet of $U(2)$.
	Only the third generation of the fermion can obtain a mass in the limit of
	unbroken symmetry limit.
  It is assumed that the quark and lepton mass matrices can be adequently
described by 
  VEV of {\it flavons} $\phi^a$,  $S^{ab}$ and $A^{ab}$, in which
   $S^{ab}$ and $A^{ab}$ are symmetric and antisymmetric tensors, respectively.
  Furthermore, it is  assumed $\langle \phi^2 \rangle/M =\e$, $\langle
S^{22}\rangle/M=\e$ 
   and $\langle A^{12} \rangle/M=\e'$, where $M$ is the cutoff scale of a
   flavon effective theory. Other VEV's are assumed to be zero.
    Thus, the $U(2)$ symmetry breaks to $U(1)$ with breaking parameter $\e$ and
	 $U(1)$  breaks to nothing with $\e'$.

  The neutrino mass matrix has been  discussed by Carone and Hall \cite{CaHa}.
  The right-handed Majorana mass  matrix
  gives a zero Majorana mass due to
   the absence of the contribution from the antisymmetric flavon  $A^{12}$.
  They proposed a simple solution which is to relax the assumptions:
 $\langle \phi^1 \rangle =0$, $\langle S^{11} \rangle =0$ and  $\langle
S^{12} \rangle =0$. 
     We present the systematic analyses of the modifed  neutrino mass matrices
  focusing on  neutrino oscillations.

 In the $SU(5)\times U(2)$ model, the modified Majorana mass matrix for the
right-handed
 neutrinos is generated at leading order by the operators
  \begin{equation}  
  \Lambda_R \left (\n_3 \n_3 +{1\o M}\phi^a \n_a\n_3+
    {1\o M^2}\phi^a\phi^b \n_a\n_b+{1\o M^3}S^{ab}\Sigma_Y\Sigma_Y\n_a\n_b
   \right ) \ ,  
    \label{Majorana} 
 \end{equation}
 \noindent where $\n_3$ and $\n_a$ are $SU(5)$ singlets, 
  and $\Sigma_Y$ is a flavor singlet and a  $\bf {24}$ of $SU(5)$.
 On the other hand, the neutrino Dirac mass matrix is generated by the operators
 \begin{equation}  
  \bar F_3 H \n_3 +{1\o M}(\phi^a\bar F_3 H\n_a+ \phi^a \bar F_a H\n_3 +
   A^{ab}\bar F_a H\n_b )+{1\o M^2}(\phi^a\phi^b \bar F_a H\n_b+
   S^{ab}\Sigma_Y\bar F_a H\n_b),   
 \end{equation}
 \noindent where $\bar F$ and $H$ are $\bar{\bf 5}$ and ${\bf 5}$
representations
  of matter and Higgs scalar, respectively.
 The $SU(5)$ representations of $\phi^a$, $S^{ab}$ and $A^{ab}$ are
assigned to be
   $\bf 1$, $\bf 75$ and $\bf 1$, respectively, in order to reproduce the 
    quark mass hierarchy.
	As far as VEV's of $\phi^1$, $S^{11}$ and  $S^{12}$ vanish, one of the
right-handed
	Majorana neutrinos is massless as seen in eq.(\ref{Majorana}).
	Therefore, we study three cases of the non-zero VEV, which were considerd 
	by Carone and Hall \cite{CaHa}:
 (1) $\langle\phi^1\rangle/M=\d_1$, (2) $\langle S^{11}\rangle/M=\d_2$ and 
 (3) $\langle S^{12}\rangle/M=\d_3$.
 The value of $\d_i$ should be ${\cal O}(\e')$ in the {\it standard} $U(2)$
model
  because there is no mechanism to provide $\d_i$ far below the $U(1)$ breaking
  parameter $\e'$.  However, we take $\d_i$ as a free parameter in the model, so
 the value of $\d_i$ is  constrained by investigating  quark masses and mixings.
 We do not address to the origin of $\d_i$ far below $\e'$ in this paper.
 We will find some patterns of  neutrino mixings depending on the magnitude
of $\d_i$.
  
 The modified neutrino mass matrices are generally written as follows:
\begin{equation}  
  M_{RR} =  
  \Lambda_R \left (
  \matrix{\d_1^2+\t\d_2\rho &  \e \d_1+\t\d_3\rho  & \d_1 \cr
\e\d_1+\t\d_3\rho & a\e^2 & \e \cr
       \d_1 & \e & 1 \cr} \right ) , \quad
  M_{LR} =  \langle H\rangle \left (
  \matrix{ \d_1^2+\t\d_2  & \e'+\t\d_3 & \d_1 \cr -\e'+\t\d_3 & \e^2 & \e \cr
       \d_1 & \e & 1 \cr} \right )  , 
	   \label{matrix}  
 \end{equation}
 \noindent
 where $\t\d_i\equiv \rho \d_i$ with $\rho\equiv \langle\Sigma_Y\rangle/M$, and
 coefficients of  ${\cal O}(1)$ are omitted except for the coefficient $a$
in each entry.
   It is noted that those coefficients should be chosen  to give the
  non-singular neutrino mass matrix.
  
  Let us discuss  following three cases with $a\not = 1$.
  The case with $a = 1$ will be discussed later.
  The light neutrino mass matrix $M_{LL}$ can be easily computed by
  the see-saw mechanism.

 {\bf case (1): $\langle \phi^1\rangle /M=\d_1\not = 0$}
  
  Taking $\d_2=0$ and $\d_3=0$ in eq.~(\ref{matrix}), we obtain
  \begin{equation}  
  M_{LL} \simeq {\langle H\rangle ^2 \o \k_1^2\Lambda_R}   
        \left (\matrix{\l^2\k_1^2 & \l\k_1 &  \d\k_1^2 \cr 
		\l\k_1 & (a-1) & \k_1 \cr
     \d\k_1^2 & \k_1 & \k_1^2 \cr} \right )  \ ,
	 \label{MLL}
 \end{equation}
 \noindent
 where we define
  \begin{equation}  
  \l \equiv {\e'\o \e}\simeq 0.22 \ , \qquad\qquad 
  \k_1 \equiv {\d_1\o \e'}  \ .
 \end{equation}
 \noindent
 Analyses of quark masses and mixings give the constraint for the magnitude
of $\k_1$.
  The serious constraint follows from  $|V_{ub}/V_{cb}|$, which is
expressed in terms of
  quark masses as follows:
    \begin{equation}  
  \left |{V_{ub}\o V_{cb}}\right | \simeq \sqrt{{m_u\o m_c}}\left (1+C\k_1
\right ) \ ,
  \end{equation}
  \noindent where $C$ is a complex coefficients of ${\cal O}(1)$.
  Taking into consideration
   the experimental values 
   $|V_{ub}/V_{cb}|=0.08\pm 0.02$ and  $\sqrt{m_u/m_c}=0.06 \pm 0.01$
\cite{PDG},
 we find  the safe constraint $\k_1 \equiv \d/\e' \leq 0.5$.

 Now, we can obtain the mixing matrix $V$, which corresponds to 
	the mixing matrix obtained in neutrino oscillation experiments, by calculating
	  $U_E^\dagger U_\n$.  
	  The unitary matrix $U_\n$ is obtained
	   by diagonalizing the matrix in eq.(\ref{MLL}), while
	  the unitary matrix $U_E$ 
	  which diagonalizes the charged lepton mass matrix $M_E$, is given as follows:
	  
  \begin{equation}  
  U_{E} \simeq  \left (\matrix{1& \sqrt{{m_e\o m_\m}} & 0 \cr
         -\sqrt{{m_e\o m_\m}} &1  & {1\o 3}{m_\m\o m_\tau}\cr
       {1\o 3}{m_\m\o m_\tau}\sqrt{{m_e\o m_\m}} & -{1\o 3}{m_\m\o m_\tau}&
1 \cr} \right )
	    \ , 	   
 \end{equation}
 \noindent
 where the $CP$ violating phase is neglected.

  The neutrino mass ratio is obtained approximately as follows:
   \begin{equation}  
  m_3 : m_2 : m_1 \simeq 1 : \k_1^2 : \l^2 \k_1^2 \ ,
  \label{ratio}
 \end{equation}
\noindent
 and the mixing matrix $V$ is given as:
 \begin{equation}  
  V \simeq \left (\matrix{1 & \l+{1\o 3}{m_\m\o m_\tau}\sqrt{{m_e\o m_\m}}
    & \l\k_1-\sqrt{{m_e\o m_\m}} \cr 
	\l\k_1 + \sqrt{{m_e\o m_\m}} & -\k_1- {1\o 3}{m_\m\o m_\tau}& 1\cr
      -\l & 1 & \k_1+{1\o 3}{m_\m\o m_\tau} \cr} \right ) \ . 
 \label{V1}
 \end{equation}
 \noindent  
 The heviest neutrino is the $\m$-like  one and
  the maximal mixing which is consistent with the atmospheric neutrino
anomaly is not obtained.
 The mixing $V_{e2}$ cannot solve the solar neutrino deficit
  because $V_{e2}\simeq \l \simeq 0.22$ is too large for
  the  small angle solution \cite{small} in the  resonant MSW transitions
\cite{MSW}:
  $\Delta m^2 =  (3 \sim 12) \times 10^{-6}~ \eV^2$ with
	 $\sin^2 2\th= (4\sim 12 )\times 10^{-3}$.
 
  Taking account of the formula of neutrino oscillations
   in the case of $m_3\gg m_2\gg m_1$:
 \begin{equation}        
   P(\n_\m\Ar \n_\a)\simeq 4 V^2_{\a 3} V^2_{\m 3} \sin^2({\Delta m^2_{31}
L \o 4 E_\n}) \ , 
   \quad (\a=e,\ \tau) \ ,
   \end{equation}
  \noindent
  we can discuss the possiblity to find neutrino oscillations in
  the accelerator and reactor neutrino experiments.
  The constraint for the heaviest neutrino mass $m_3$ depends on the mixing
  $|V_{e3}|\simeq |\l\k_1-0.07|$.
  As far as $\k_1=0.20\sim 0.44$ in eq.(\ref{V1}), 
  the experimental upper bound of $P(\n_\m\Ar \n_e)$ \cite{Bugey,E776} 
   allows the neutrino mass of  $m_3\geq 1 ~\eV$,
   which is consistent with  the hot dark matter (HDM).
  Moreover, this mixing matrix is also consistent with  the LSND data, which
   suggest  typical parameters such as  $\Delta m^2 \sim 2~\eV^2$
    with $\sin^2 2\th_{\rm LSND}  \simeq 2\times 10^{-3}$ \cite{LSND}. 
 Then, in the  CHORUS and  NOMAD experiments \cite{CHONOM},
  we predict  $P(\n_\m\Ar \n_\tau)= (1\sim 4)\times 10^{-4}$, which may be a 
  detectable magnitude.
  
  LBL accelerator experiments are planed to operate
   in the near future \cite{long,KEKPS}.
   The first experiment will begin  in K2K~(250~Km).
  For LBL experiments, the relevant formula of neutrino oscillations are
 \begin{equation}        
 P(\n_\m\Ar \n_\a) \simeq -4 V_{\a 1} V_{\m 1} V_{\a 2} V_{\m 2} 
            \sin^2({\Delta m^2_{21} L \o 4 E_\n})+
            2 V^2_{\a 3} V^2_{\m 3}  \ , \quad (\a=e,\ \tau) \ ,
\end{equation}
\noindent where we assume  $\Delta m^2_{31}\gg\Delta m^2_{21}$.
 Let us predict the neutrino oscillation  in K2K, where  the average energy
of the $\n_\m$ beam 
    is taken as  $E_\n=1.3~\G$ with $L=250$~Km.
	 If  we take $\k_1=0.2\sim 0.4$ and $\Delta m_{21}^2=0.01~\eV^2$ tentatively,  
	  we predict
	    $P(\n_\m\Ar \n_e)=0.01\sim 0.03$ and $P(\n_\m\Ar \n_\tau)=0.1\sim 0.4$.
	Thus, the K2K experiment is expected to detect
	   signatures of  neutrino oscillations in this model.

{\bf case (2): $\langle S^{11}\rangle /M=\d_2\not = 0$}
 
Taking $\d_1=0$ and $\d_3=0$ in eq.~(\ref{matrix}), we obtain
the light neutrino mass matrix:
 
  \begin{equation}  
  M_{LL} \simeq {\langle H\rangle ^2 \o (a-1)\rho \Lambda_R}   
        \left (\matrix{\rho \l^2 & -(a-1)\e' & \rho\l \cr 
		-(a-1) \e' & (a-1)\e'^2/\t \d_2 & \e \rho \cr
       \rho\l & \e \rho & \rho \cr} \right ) \ .
 \end{equation}	
 We get the mass ratio as:
   \begin{equation}  
  m_3 : m_2 : m_1 \simeq 1 : \t\k_2/\l : \l \t\k_2 \ ,
 \end{equation}
 \noindent
   where $\t\k_2\equiv \t\d_2 /\e'$ is defined. The constrait for  $\t\k_2$
   is given by  $V_{us}$, which is expressed in terms of
  quark masses as follows:
    \begin{equation}  
  |V_{us}|\simeq \sqrt{{m_d\o m_s}}\left (1+{\k_2\o\l}\right )^{-{1\o 2}}
   -\sqrt{{m_u\o m_c}}\left (1+C{\k_2\o\l}\right )^{-{1\o 2}} \ .
  \end{equation}
  \noindent 
    Taking into consideration the accuracy of $5\%$ for
$|V_{us}|=\sqrt{{m_d/ m_s}}$,
  we obtain a tight constraint $\k_2  \leq 0.1 \l\simeq 0.02$, 
  which gives  $\t\k_2\leq 0.02\rho\simeq 0.0004$.
	 The mixing matrix $V$ is given as
 \begin{equation}  
  V \simeq \left (\matrix{1 & \l+{1\o 3}{m_\m\o m_\tau}\sqrt{{m_e\o m_\m}}
    & -\t\k_2-\sqrt{{m_e\o m_\m}} \cr 
	\t\k_2 + \sqrt{{m_e\o m_\m}} & \l\t\k_2-{1\o 3}{m_\m\o m_\tau}& 1\cr
      -\l & 1 & -\t\k_2^2+{1\o 3}{m_\m\o m_\tau} \cr} \right ) \ .  
 \end{equation}
 \noindent 
 In this case,  the heviest neutrino is also the  $\m$-like one  and
  the large mixing which is consistent with the atmospheric neutrino
anomaly cannot be 
   realized.
 Since $\t\k_2$ is very small, we cannot find  new interesting phenomena.
 This mixing matrix also cannot solve the solar neutrino problem
    due to  $V_{e2}\simeq \l \simeq 0.22$.
 
  If the heaviest mass $m_3$ is constrained to be $m_3\leq 0.8~\eV$,
  the LSND data is consistent with our obtained mixing  $|V_{\e3}|\simeq 0.07$.
   However, this mass value is not consistent with  HDM. 
  In the  LBL experiment at K2K, 
   we can expect $P(\n_\m\Ar \n_e)\simeq 0.01$ and $P(\n_\m\Ar
\n_\tau)\simeq 0.01$
   as far as $m_3\geq 0.1~\eV$ is statisfied.

{\bf case (3): $\langle S^{12}\rangle /M=\d_3\not = 0$}
 
Taking $\d_1=0$ and $\d_2=0$ in eq.~(\ref{matrix}), 
the light neutrino mass matrix is given as:
 \begin{equation}  
  M_{LL} \simeq {\langle H\rangle ^2 \o \rho^2\t\d_3^2 \Lambda_R}   
        \left (\matrix{(1-a)\e^2\t\d_3^2 & (a-1)\e^2\e'\t\d_3 &
\rho\e\t\d_3^2 \cr 
		 (a-1)\e^2\e'\t\d_3 & (1-a)\e^2\e'^2 & \e\e' \rho\t\d_3 \cr
       \rho\e\t\d_3^2 & \e\e' \rho\t\d_3   &  \rho^2\t\d_3^2\cr} \right ) \ .
	   \label{MLL3} 
 \end{equation}	
   The hierarchy of this matrix is somewhat complicated, but using the relation
     $\e\simeq \rho$ we obtain approximately the mass ratio:
   \begin{equation}  
  m_3 : m_2 : m_1 \simeq 1 : \left |{a\o 1-a}\right |\t\k_3^2 :
     \left |{1\o a(1-a)}\right |\t\k_3^2 \ ,
	 \label{massra}
 \end{equation}
 \noindent where $\t\k_3$ is constrained in the quark sector.
 Since the $|V_{us}|$ is expressed in terms of quark masses as follows:
    \begin{equation}  
  |V_{us}|\simeq \sqrt{{m_d\o m_s}}(1+\k_3)
   -\sqrt{{m_u\o m_c}} (1+\k_3)   \  ,
  \end{equation}
  \noindent
  we obtain a tight constraint  $\k_3  \leq 0.05$,
  which leads to $\t\k_3\leq 0.05\rho\simeq 0.001$.
 Therefore, we predict  the  huge mass hierarchy ${\cal O}(10^{6})$ in
eq.(\ref{massra}).
 
 The mixing matrix $V$ is obtained as:
 \begin{equation}  
  V \simeq \left (\matrix{1 & \a+{1\o 3}{m_\m\o m_\tau}\sqrt{{m_e\o m_\m}}
    & -\t\k_3-\sqrt{{m_e\o m_\m}} \cr 
	\t\k_3 + \sqrt{{m_e\o m_\m}}+\a{1\o 3}{m_\m\o m_\tau} & 
	        \a\t\k_3+\a\sqrt{{m_e\o m_\m}}-{1\o 3}{m_\m\o m_\tau}& 1\cr
     -\a & 1 & \t\k_3+{1\o 3}{m_\m\o m_\tau} \cr} \right ) \ ,  
 \end{equation}
 \noindent where $\a$ is a coefficients of ${\cal O}(\leq\l)$.
 This mixing matrix may  solve the solar neutrino problem
   since $V_{e2}\simeq \a$ could be small.
   However, $m_3$ should be larger than $10^3~\eV$ as seen in eq.(\ref{massra})
    if the solar neutrino mass scale 
	$\Delta m_{21}^2 \simeq 10^{-6}~\eV^2$ is fixed.  This large mass $m_3$ is
not consistent
	with the present experimental bound of E776 \cite{E776} 
	because of $|V_{e3}|\simeq 0.07$.
	In conclusion, the solar neutrino problem is not explained 
	 as well as the atmospheric neutrino deficit in this case.
	On the other hand, for the  LBL experiment at K2K, 
   we predict $P(\n_\m\Ar \n_e)\simeq 0.01$ and $P(\n_\m\Ar \n_\tau)\simeq
0.01$.
  

Let us consider a possibility of the  maximal mixing
 which is reconciled with
 the atmospheric neutrino anomaly.
 If $a=1$ is satisfied exactly with  $k_1^2\ll 1$ in eq.(\ref{MLL}) of case (1)
  or with  $\t\d_3\ll 1$ in eq.(\ref{MLL3}) of  case (3), 
  this matrix gives the maximal mixing
  because the diagonal entries of the matrix  are suppressed.
   Then, the mixing matrix with the maximal mixing is given for both cases
as follows:
  \begin{equation}
   V  \simeq   {1\o \sqrt{2}}\left (\matrix{\sqrt{2} &  
    \l-\sqrt{{m_e\o m_\m}}+{1\o 3}{m_\m\o m_\tau}\sqrt{{m_e\o m_\m}} & 
	\l +\sqrt{{m_e\o m_\m}}+{1\o 3}{m_\m\o m_\tau}\sqrt{{m_e\o m_\m}}\cr 
		\sqrt{2}\sqrt{{m_e\o m_\m}} & 1 & -1 \cr  -\sqrt{2} \l & 1 & 1  \cr}
\right ) \ .
  \label{max}
 \end{equation}
 \noindent  
 The mass ratio is approximately given for case (1) (case(3)):
   \begin{equation}  
  m_3 : m_2 : m_1 \simeq 1 : 1 : \l^2 \k_1 (\l^2 \t\k_3) \ .
 \end{equation}
 Actually, for the case (1), 
 the choice of $\k_1 \simeq 0.01$ reproduces the maximal $\nu_{\mu} -
\nu_{\tau}$ mixing
 with \cite{Atm2,Kam}
 \begin{equation}
\Delta m_{32}^2  =  (0.3 \sim 3) \times 10^{-2}~ \eV^2 \  .    
\label{atm}
\end{equation}
 \noindent
 Therefore, the LBL experiment at K2K will find
  the considerably large $\nu_{\mu} - \nu_{\tau}$ oscillation probability.
  For the $\nu_{\mu} - \nu_e$ oscillation, $P(\n_\m\Ar \n_e)\simeq 0.03$ is
expected.
  
 What happens for the accelerator and reactor neutrino experiments in this case?
   Since $m_3\simeq m_2\gg m_1$ is satisfied,
  the relevant formula of neutrino oscillations are given as:
 \begin{equation}        
   P(\n_\m\Ar \n_\a)\simeq 4 V^2_{\a 1} V^2_{\m 1} \sin^2({\Delta m^2_{31}
L \o 4 E_\n}) \ , 
   \quad (\a=e,\ \tau) \ .
 \end{equation}
  \noindent 
  The accelerator and reactor neutrino experiment of $\n_\m\Ar \n_e$
\cite{Bugey,E776,LSND}
  strongly constrains the value of the heaviest neutrino mass $m_3$ 
   since $V_{\m1}\simeq 0.07$ is considerbly large as already discussed by
   Carone and Hall \cite{CaHa}.
 In this case, one obtains $m_3\leq 0.8~ \eV$ \cite{Bugey,E776}, which is
not consistent with HDM.
  If the LSND data is taken seriously, $m_3=0.5\sim 0.8 ~\eV$ is obtained.
 Then, in the  CHORUS and  NOMAD experiments \cite{CHONOM},
  we predict  $P(\n_\m\Ar \n_\tau)\simeq 4\times 10^{-7}$ and
  $P(\n_\m\Ar \n_e)\simeq 8\times 10^{-5}$, which are out of 
  their experimental sensitivity.

 For the case (3),
 the choice of $\t\k_3 \simeq 0.001$ explains easily the atmospheric
neutrino deficit 
  by reproducing $\Delta m_{32}^2 \simeq 10^{-2}~ \eV^2 $     
 with the $\nu_{\mu} - \nu_{\tau}$ oscillation.
 Predictions of neutrino oscillations are  same ones
  as the ones in  case (1).
 
   Since above results are very interesting, it may be important to give
following comments.
  The very small value of $\k_1$ and $\t\k_3$  cannot be understood
   solely in terms of a $U(2)$ symmetry breaking pattern.
 Moreover, the $U(2)$ flavor symmetry does not guarantee the value of  $a=1$.
 
 For above all cases,  we summarize
    expected signatures of neutrino oscillation experiments  in Table 1.
 The model cannot solve the solar neutrino deficit while
  it could be reconciled with the  LSND data
   by taking the adjustable $m_3$.
  In the special case of $a=1$ in the right-handed 
 Majorana mass matrix, the maximal mixing of $\nu_{\mu} - \nu_{\tau}$ is
derived. Therefore, 
  the atmospheric neutrino anomaly is explained by the large 
  $\nu_{\mu} - \nu_{\tau}$ oscillations in those cases.
  Thus, we have clarified the
   phenomenological implications of the $SU(5)\times U(2)$  model
    in the neutrino sector.
	The model will be tested seriously in comming neutrino oscillation
	 experiments.

I would like to thank  Prof. M. Matsuda for the  helpful
 discussions and careful reading the manuscript. 

\newpage


\newpage

\topskip 1 cm
\begin{table}
\hskip 1 cm
\begin{tabular}{|l |l |l |l |} \hline
           &                   &          &                  \\
 \quad  Cases & \quad $\langle \phi^1\rangle /M=\d_1$  & 
        \  $\langle S^{11}\rangle /M=\d_2$  \ &  
         \quad $\langle S^{12}\rangle /M=\d_3$  \\
		   &            &          &        \\
         &  $a\not=1$ \qquad \   $a=1$  &      &  $a\not=1$ \qquad \   
$a=1$   \\ \hline
               &               &          &        \\
  Solar    &   No\  \qquad\quad  No    & \qquad No    & No\ \qquad\quad  No  \\
     &            &          &        \\
  Atmospheric     &  No\   \qquad\quad Yes      & \qquad No    & No\   
\qquad\quad Yes  \\
     &            &          &        \\
  LSND      & Yes  \qquad\quad Yes  &  \qquad Yes  & Yes\qquad\quad Yes\\
     &            &          &        \\
  HDM      &   Yes \qquad\quad No  &  \qquad No  &  No\ \qquad\quad No \\
     &            &          &        \\
  CHORUS/      & Yes \qquad\quad No    & \qquad No  &  No\ \qquad\quad No\\
  NOMAD  &            &          &        \\
      &            &          &        \\
  LBL(K2K)    & Yes \qquad\quad Yes  & \qquad Yes   &   Yes\qquad\quad Yes \\
                 &            &          &     \\  \hline
\end{tabular}
\caption{Expected experimental signatures of  neutrino oscillations in the
model.
  "Yes" denotes that signatures are detectable if the relevant neutrino masses
   are taken.   On the other hand, "No" denotes that
      the model cannot explain  the phenomenon. }
\end{table}

\end{document}